\documentclass[fp,twocolumn]{jpsj3}
\usepackage{txfonts}
\usepackage{bm}

\title{Two-Step Discontinuous Shear Thickening of Dilute Inertial Suspensions Having Soft-Core Potential}

\author{Shuichi Sugimoto$^1$ and Satoshi Takada$^{1,2}$\thanks{Corresponding Author, takada@go.tuat.ac.jp}}
\inst{
	$^1$Department of Mechanical Systems Engineering, Tokyo University of Agriculture and Technology, 2--24--16, Naka-cho, Koganei, Tokyo 184--8588, Japan \\
	$^2$Institute of Engineering, Tokyo University of Agriculture and Technology, 2--24--16, Naka-cho, Koganei, Tokyo 184--8588, Japan
} 

\abst{
Kinetic theory for dilute inertial suspension having soft-core potential is theoretically investigated.
From the analysis of the scattering process, the expression of the scattering angle is analytically obtained.
We derive the flow curve between the viscosity and the shear rate, which shows two-step discontinuous shear thickening when we change the softness of the particles.
The molecular dynamics simulation shows that our theoretical results are consistent with the numerical ones.
}
\begin{document}
\maketitle

%%%%%%%%%%%%%%%%%%%%%%%%%%%%%%
\section{Introduction}
It is important in many situations to understand how the system flows.
The hydrodynamic treatment is effective to know the rheological properties of the system because theoretical analysis may be available from the continuum description, for example, the hydrodynamic equations.
For this purpose, we should know the expressions of the transport coefficients for this system, such as the shear viscosity or the thermal conductivity.
One of the most powerful tools to understand these coefficients is the kinetic theory.
The kinetic theory for dilute hard-sphere gases has been well studied after Boltzmann \cite{Chapman}.
We know that the explicit expressions of them from the Boltzmann equation when we adopt the Chapman-Enskog theory \cite{Chapman, Ferziger}.
Many papers studied the validity of them using simulations such as the molecular dynamics (MD) simulation or the direct simulation Monte Carlo method.
The denser systems are also studied using the Enskog theory, where the size of the particles should be considered \cite{Resibois}.
Some papers have studied the rheology of the inertial suspension of hard-sphere particles, which is a kind of an idealistic setup of aerosols, in terms of the kinetic theory\cite{Tsao95, Sangani96, Hayakawa19, Hayakawa17, Saha20, Takada20}, where the theory can predict discontinuous shear thickening (DST) for dilute situations.
Here, the origin of a DST-like process in this system is a bifurcation of quenched-ignited transition \cite{Tsao95}, and this is different from denser systems, which is the transition between liquid-like and solid-like phases \cite{Seto13, Fernandez13, Mari15, Guy15, Hsiao17, Kawasaki18}.

However, the assumption of the hard-core potential is more or less idealistic, because the actual particles deform when colliding with each other.
The simplest model for deformable particles is the Hookean, which means that the repulsive force between the deformed particles is proportional to the overlap length between particles.
This model is also known as the harmonic potential, which can describe the deformation when it is sufficiently small.
We sometimes use this model to perform the MD simulations due to its simple treatment of collisions.
The results obtained from this potential deviate from those with hard-core limit when the mean velocity of the system becomes larger.
This is because the overlap between particles is finite, which is not considered in the hard-core limit.

Although its simpleness, the suspension model with this potential exhibits complex rheology \cite{Berthier09, Kawasaki14, Philippe18}.
Kawasaki {\it et al.} \cite{Kawasaki14} used this model for denser cases, and they reported that the flow curve shows the shear thinning, shear thickening, and again shear thinning behaviors as the shear rate increases when the density is lower than the jamming density.
This model is also known to show a divergent behavior of the relaxation time \cite{Philippe18}.
To understand these behaviors step by step, the theoretical treatment for dilute systems must be its first step.

To this end, we expand the kinetic theory to the homogeneous system consisting of the particles having the soft interparticle potential.
We note that the kinetic theory works well even for the system having the Lennard-Jones potential \cite{Hirschfelder, Kihara43}, the square-well potential \cite{Holleran51, Hirschfelder, Takada16, Takada18}, the penetrable square well potential \cite{Sanchez19}, or other potentials \cite{Hirschfelder}.
Thus, we expect that the kinetic theory is also applicable to this system.
In this paper, we adopt the harmonic potential as an intermolecular potential.
As far as we know, there are no papers studying the transport coefficients of this system.
But as discussed later, the scattering process of this system is analytic, and we can obtain the explicit expression of the scattering angle as a function of the impact parameter and the relative speed, which is the advantage to use this potential.
Once we numerically calculate the integral which characterizes the model, we can easily calculate the transport coefficients by solving a set of equations which determine the rheology of the system.

The organization of this paper is as follows:
In the next section, we develop the kinetic theory of the dilute gas-solid suspension system.
In Sec.\ \ref{sec:scattering}, the scattering angle is derived.
In Secs.\ \ref{sec:kinetic_theory} and \ref{sec:rheology}, we briefly explain the development of the kinetic theory to our system and derive the microscopic expressions of the shear viscosity and the other quantities.
We also perform the MD simulation to validate our theory in Sec.\ \ref{sec:MD}.
In the last two sections, we discuss and conclude our results.
We also have two Appendixes.
In Appendix \ref{sec:quintic}, we briefly explain the procedure to solve the quintic equation appeared when we solve the scattering process.
In Appendix \ref{sec:theta}, the expression of the scattering angle is analytically derived using the solution obtained in Appendix \ref{sec:quintic}.

%%%%%%%%%%%%%%%%%%%%%%%%%%%%%%
\section{Model}\label{sec:model}
\begin{figure}[htbp]
	\centering
	\includegraphics[width=\linewidth]{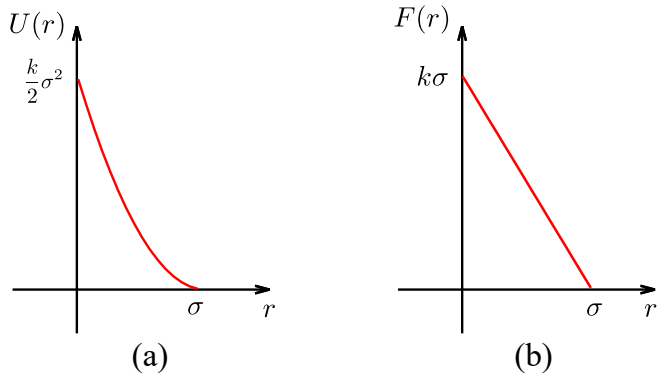}
	\caption{(Color online) The schematic pictures of (a) the potential and (b) the force between particles.}
	\label{fig:potential_force}
\end{figure}

We consider the monodisperse particles, whose mass and diameter are given by $m$ and $\sigma$, respectively, in the three-dimensional system.
Here, we assume that the particles interact with each other via the harmonic potential
\begin{equation}
	U(r_{ij})=\frac{k}{2}(\sigma-r_{ij})^2\Theta(\sigma-r_{ij}),\label{eq:potential}
\end{equation}
where, $r_{ij}=|\bm{r}_{ij}|=|\bm{r}_i-\bm{r}_j|$ is the distance between $i$-th and $j$-th particles, $k$ represents the strength of the repulsive force, and $\Theta(x)$ is the step function.
This potential \eqref{eq:potential} means that the repulsive force between particles is the Hookean $F(r_{ij})=-\partial U(r_{ij})/\partial r_{ij}=k(\sigma-r_{ij})\Theta(\sigma-r_{ij})$.
The schematic pictures of the potential and the force are shown in Fig.\ \ref{fig:potential_force}.

%%%%%%%%%%%%%%%%%%%%%%%%%%%%%%
%%%%%%%%%%%%%%%%%%%%%%%%%%%%%%

%%%%%%%%%%%%%%%%%%%%%%%%%%%%%%
\section{Scattering Process}\label{sec:scattering}
\begin{figure}[htbp]
	\centering
	\includegraphics[width=0.75\linewidth]{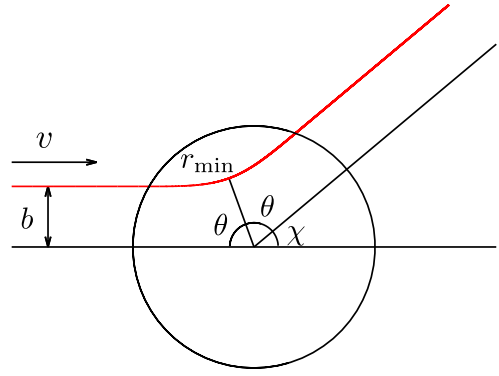}
	\caption{(Color online) The schematic picture of the scattering process, where $v$ is the relative speed, $b$ is the collision parameter, and $r_{\rm min}$ is the closest distance.}
	\label{fig:scattering}
\end{figure}

First, let us study the scattering process as shown in Fig.\ \ref{fig:scattering}.
When two particles approach each other with the impact parameter $b$ and the relative speed $v$, the angle between the infinity and the closest position is known to be given by
\begin{equation}
	\theta=\int_0^{u_0}\frac{b du}{\sqrt{1-b^2u^2-\frac{4}{mv^2}U(\frac{1}{u})}},\label{eq:theta_def}
\end{equation}
where $u_0(>1/\sigma)$ is the root of the denominator of the integrand \cite{Goldstein}.

%%%%%%%%%%%%%%%%%%%%%%%%%%%%%%
%%%%%%%%%%%%%%%%%%%%%%%%%%%%%%

For $b>\sigma$, there happen no collision, which means $u_0=1/b$ and the angle becomes
\begin{equation}
	\theta=\int_0^{1/b}\frac{bdu}{\sqrt{1-b^2u^2}}=\frac{\pi}{2}.
\end{equation}

For $b\le d$, on the other hand, Eq.\ \eqref{eq:theta_def} is rewritten as
\begin{align}
	\theta&=\int_0^{1/\sigma}\frac{bdu}{\sqrt{1-b^2u^2}}
	+\int_{1/\sigma}^{u_0}\frac{bdu}{\sqrt{1-b^2u^2-\frac{2k}{mv^2}(\sigma-\frac{1}{u})^2}}\nonumber\\
	&= \sin^{-1}\frac{b}{\sigma}+\int_{1}^{u_0^*}\frac{u^* du^*}{\sqrt{-(u^{*4}+p u^{*2}+qu^*+r)}},\label{eq:theta_eq1}
\end{align}
where we have introduce the dimensionless quantities $b^*\equiv b/\sigma$, $u^*\equiv u\sigma$, $u_0^*\equiv u_0\sigma$, and 
\begin{equation}
	p\equiv \frac{2-v^{*2}}{b^{*2}v^{*2}},\quad
	q\equiv -\frac{4}{b^{*2}v^{*2}},\quad
	r\equiv \frac{2}{b^{*2}v^{*2}}=-\frac{q}{2}.
	\label{eq:def_pqr}
\end{equation}
For further treatment, we also introduce the following parameters:
\begin{align}
	P&\equiv -\left(\frac{p^2}{3}+4r\right),\quad
	Q\equiv -\frac{2}{27}p^3-q^2+\frac{8}{3}pr,\\
	\Delta &\equiv \left(\frac{Q}{2}\right)^2+\left(\frac{P}{3}\right)^3.\label{eq:Delta}
\end{align}
Here, $u_0^*$ is a solution of the quartic equation $u^{*4}+p u^{*2}+qu^*+r=0$, which should satisfy $u_0^*\ge 1$.
We note that this equation is always solvable by Ferrari's method.
First, let us define $\beta$ as
\begin{equation}
	\beta \equiv 
	\begin{cases}
		-\frac{2p}{3}+\left(-\frac{Q}{2}+\sqrt{\Delta}\right)^{1/3}+\left(-\frac{Q}{2}-\sqrt{\Delta}\right)^{1/3}
		& (\Delta \ge0)\\
		-\frac{2p}{3}+3\sqrt{-\frac{P}{3}} \cos\left\{\frac{1}{3}\cos^{-1}\left[-\frac{Q}{2}\left(-\frac{3}{P}\right)^{3/2}\right]\right\}
		& (\Delta<0)
	\end{cases}.
	\label{eq:beta}
\end{equation}
Here, it is noted that the condition $\Delta\ge0$ is equivalent to the following condition:
\begin{align}
	&(0\le v^*\le \sqrt{2},\ 0\le b^*\le 1)\ \cup \nonumber\\
	&\left(\sqrt{2}<v^*\le \sqrt{11+5\sqrt{5}},\ b_{\rm min}^*\le b^*\le 1\right),
	\label{eq:v_cond}
\end{align}
with
\begin{equation}
	b_{\rm min}^*\equiv \frac{\sqrt{2v^{*4}+10v^{*2}-1-(4v^{*2}+1)^{3/2}}}{4v^*}.
\end{equation}
Using this $\beta$, we can write explicitly $u_0^*$ as
\begin{equation}
	u_0^*= \frac{\sqrt{\beta} + \sqrt{-\beta-2p - \frac{2q}{\sqrt{\beta}}}}{2}.\label{eq:u0}
\end{equation}

%%%%%%%%%%%%%%%%%%%%%%%%%%%%%%
\begin{figure}[htbp]
	\centering
	\includegraphics[width=\linewidth]{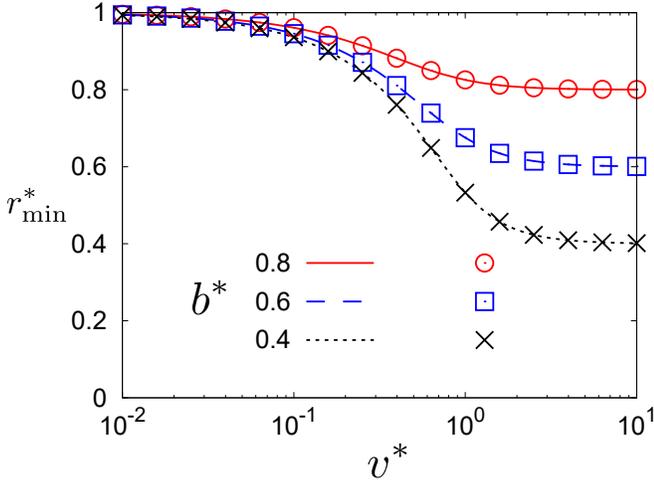}
	\caption{(Color online) Plots of the dimensionless closest distance $r_{\rm min}^*(\equiv r_{\rm min}/\sigma)$ as a function of the dimensionless relative speed $v^*(\equiv v/(\sigma\sqrt{k/m}))$ for $b^*(\equiv b/\sigma)=0.8$ (red solid line), $0.6$ (blue dashed line), and $0.4$ (black dotted line).
	The simulation results are also plotted for the same $b^*$.}
	\label{fig:r_min}
\end{figure}
%%%%%%%%%%%%%%%%%%%%%%%%%%%%%%

Figure \ref{fig:r_min} shows the velocity dependence of $r_{\rm min}=1/u_0$.
The magnitude of the deformation ($\sigma-r_{\rm min}$) is small for the low-speed regime, and it increases as the relative speed increases, then converges to the impact parameter $b$.
This behavior is quite natural:
For the low-speed regime, because the kinetic energy is much smaller than the potential energy, the particle reflects soon after entering the potential region, which means that the closest distance $r_{\rm min}$ is almost equal to the diameter of the particles.
On the other hand, the particle cannot feel the potential for the high-speed regime, which is the reason why the closest distance $r_{\rm min}$ converges to the impact parameter $b$.
The dependence of the closest distance on the impact parameter and the relative speed is also validated by the molecular dynamics simulations, which will be explained in the next section.

%%%%%%%%%%%%%%%%%%%%
\begin{table}[htbp]
	\centering
	\caption{The expressions of the coefficients appeared in Eq.\ \eqref{eq:theta}.}
 	\begin{tabular}{|c|c|c|}
 	\hline
 	 & condition I & condition II \\ 
 	 & $\displaystyle \beta+2p\le \frac{2q}{\sqrt{\beta}}$ & $\displaystyle \beta+2p > \frac{2q}{\sqrt{\beta}}$ \\ \hline
 	$D_1$ & \multicolumn{2}{|c|}{$\displaystyle -\beta-2p+\frac{2q}{\sqrt{\beta}}$} \\ 
 	$D_2$ & \multicolumn{2}{|c|}{$\displaystyle -\beta-2p-\frac{2q}{\sqrt{\beta}}$} \\ 
 	$\alpha_1$ & \multicolumn{2}{|c|}{$\displaystyle \frac{q+\sqrt{q^2+2\beta^2(p+\beta)}}{2\beta}$} \\ 
 	$\alpha_2$ & \multicolumn{2}{|c|}{$\displaystyle \frac{q-\sqrt{q^2+2\beta^2(p+\beta)}}{2\beta}$} \\ 
 	$A_1$ & \multicolumn{2}{|c|}{$\displaystyle \sqrt{-\frac{\beta^2 D_1}{\sqrt{q^2+2\beta^2(p+\beta)}-q-\beta^{3/2}}}$} \\ 
 	$A_2$ & \multicolumn{2}{|c|}{$\displaystyle \sqrt{\frac{\beta^2 D_2}{\sqrt{q^2+2\beta^2(p+\beta)}-q+\beta^{3/2}}}$} \\ 
 	$w_0$ & \multicolumn{2}{|c|}{$\displaystyle -\frac{\sqrt{q^2+2\beta^2(p+\beta)}+q-2\beta}{\sqrt{q^2+2\beta^2(p+\beta)}-q+2\beta}$} \\ \hline
 	$\nu$ & $\displaystyle \frac{A_2^2}{A_1^2}$ & $\displaystyle \frac{A_2^2}{A_1^2+A_2^2}$ \\ 
 	$\phi$ & $\displaystyle \sin^{-1}\frac{w_0}{A_2}$ & $\displaystyle \cos^{-1}\frac{w_0}{A_2}$ \\ 
 	$\gamma$ & $\displaystyle \sqrt{\frac{(A_1^2-1)(A_2^2-w_0^2)}{(1-A_2^2)(A_1^2-w_0^2)}}$ & $\displaystyle \sqrt{\frac{(A_1^2+1)(A_2^2-w_0^2)}{(1-A_2^2)(A_1^2+w_0^2)}}$ \\ 
 	$a$ & $\displaystyle A_2^2$ & $\displaystyle -\frac{A_2^2}{1-A_2^2}$ \\ 
 	$C$ & $\displaystyle \frac{2\sqrt{A_1A_2}}{(D_1D_2)^{1/4}}$ & $\displaystyle \frac{2\sqrt{A_1A_2}}{(-D_1D_2)^{1/4}}$ \\ 
 	$C_1^\prime$ & $\displaystyle -\frac{\alpha_2}{A_1}$ & $\displaystyle \frac{\alpha_2}{\sqrt{A_1^2+A_2^2}}$ \\ 
 	$C_2^\prime$ & $\displaystyle -\frac{\alpha_1-\alpha_2}{A_1}$ & $\displaystyle \frac{\alpha_1-\alpha_2}{(1-A_2^2)\sqrt{A_1^2+A_2^2}}$ \\ 
 	$C_3^\prime$ & $\displaystyle \frac{\alpha_1-\alpha_2}{\sqrt{(A_1^2-1)(1-A_2^2)}}$ & $\displaystyle \frac{\alpha_1-\alpha_2}{\sqrt{(A_1^2+1)(1-A_2^2)}}$ \\ 
 	$C_4^\prime$ & $\displaystyle -C_1^\prime K(\nu) - C_2^\prime \Pi(a,\nu)$ & $0$ \\ \hline
 	$C_1$ & \multicolumn{2}{|c|}{$C C_1^\prime$} \\ 
 	$C_2$ & \multicolumn{2}{|c|}{$C C_2^\prime$} \\ 
 	$C_3$ & \multicolumn{2}{|c|}{$C C_3^\prime$} \\ 
 	$C_4$ & \multicolumn{2}{|c|}{$C C_4^\prime$} \\ 
 	\hline
	\end{tabular}
	\label{fig:coeff_list}
\end{table}
%%%%%%%%%%%%%%%%%%%%
%%%%%%%%%%%%%%%%%%%%%%%%%%%%%%
\begin{figure}[htbp]
	\centering
	\includegraphics[width=\linewidth]{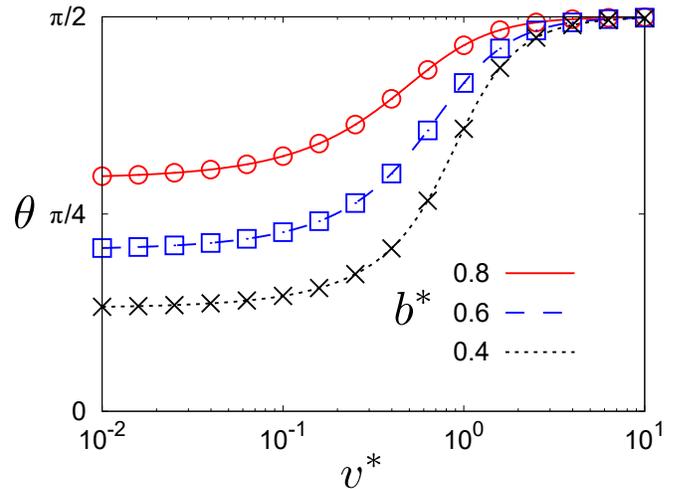}
	\caption{(Color online) Velocity dependence of the angle $\theta$ from Eq.\ \eqref{eq:theta} for $b^*=0.8$ (red solid line), $0.6$ (blue dashed line), and $0.4$ (black dotted line).
	The simulation results are also plotted for the same $b^*$.}
	\label{fig:theta}
\end{figure}
%%%%%%%%%%%%%%%%%%%%%%%%%%%%%%

Using Eq.\ \eqref{eq:u0}, we can finally obtain the expression of $\theta$ from Eq.\ \eqref{eq:theta_eq1} as
\begin{equation}
	\theta = \sin^{-1}\frac{b}{\sigma} + C_1 F(\phi,\nu) +C_2\Pi(a; \phi | \nu) +C_3\tan^{-1}\gamma + C_4,\label{eq:theta}
\end{equation}
where the expressions of $C_1$, $C_2$, $C_3$, $\phi$, $\gamma$, $\nu$, and $a$ are listed in Table \ref{fig:coeff_list}.
(The detailed derivation is given in Appendix \ref{sec:theta}.)
It should be noted that the expressions of the coefficients shown in Table \ref{fig:coeff_list} change depending on the sign of $\beta+2p-2q/\sqrt{\beta}$.
Here, $F(\phi,\nu)\equiv \int_0^{\phi}d\phi^\prime /\sqrt{1-\nu\sin^2\phi^\prime}$ and $\Pi(a,\phi|\nu)\equiv \int_0^{\phi}d\phi^\prime /\{(1-a\sin^2\phi^\prime)\sqrt{1-\nu\sin^2\phi^\prime}\}$ are the elliptic integrals of the first and third kind, respectively \cite{Abramowitz}.
We also note that we also use the complete elliptic integrals of the first and third kind as $K(\nu)\equiv F(\pi/2,\nu)$ and $\Pi(a,\nu)\equiv \Pi(a,\pi/2|\nu)$, respectively, in Table \ref{fig:coeff_list}.
For later usage, we also define the scattering angle $\chi$ (see also Fig.\ \ref{fig:scattering}) as
\begin{equation}
	\chi = \pi - 2\theta.\label{eq:chi}
\end{equation}
Figure \ref{fig:theta} shows the velocity dependence of the angle $\theta$.
The low speed limit converges to $\sin^{-1}(b/\sigma)$, which is consistent with the behavior of $r_{\rm min}$ in Fig.\ \ref{fig:r_min}.
This is because the overlap is small, and the collision is similar to that for hard-core limit.
On the other hand, the angle $\theta$ converges to $\pi/2$ in the high speed limit.
This is consistent with the fact that the trajectory of the particle is almost straightforward because the kinetic energy is much larger than the potential.
We also note that this behavior is also validated by the molecular dynamics simulations as well as the case for the closest distance.

\begin{figure}[htbp]
	\centering
	\includegraphics[width=\linewidth]{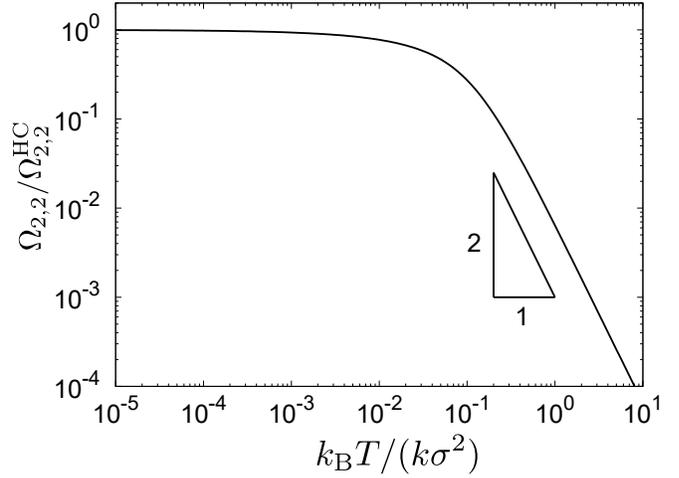}
	\caption{The temperature dependence of the Omega integral $\Omega_{2,2}$, where $\Omega_{2,2}^{\rm HC}$ represents that for hard-core gases given by $\Omega_{2,2}^{\rm HC}=2\sigma^2 \sqrt{\pi k_{\rm B}T/m}$.
	The inner triangle shows the slope of $\Omega_{2,2}$ in the high temperature regime.}
	\label{fig:Omega22}
\end{figure}

We also introduce the Omega integral $\Omega_{k,l}(T)$ \cite{Chapman, Hirschfelder, Sanchez19} as
\begin{equation}
	\Omega_{k,l}(T)= \sqrt{\frac{k_{\rm B}T}{\pi m}}\int_0^\infty dy e^{-y^2}y^{2k+3} Q_l \left(2y\sqrt{\frac{k_{\rm B}T}{m}}\right),
	\label{eq:Omega_integral}
\end{equation}
with 
\begin{equation}
	Q_l(v)=2\pi \int_0^\infty db \hspace{0.2em}b \left[1-\cos^l \chi(b,v)\right],\label{eq:Q_l}
\end{equation}
and the Boltzmann constant $k_{\rm B}$.
It is well known that $\Omega_{1,1}$ and $\Omega_{2,2}$ relate to the self-diffusion coefficient and the shear viscosity, respectively.
As discussed later, we focus on $\Omega_{2,2}(T)$ in this paper.
Because it is not possible to analytically evaluate $\Omega_{2,2}(T)$, we numerically evaluate this quantity by solving the double integral with respect to $y$ and $b$ in Eq.\ \eqref{eq:Omega_integral}.
Here, it is noted that, in the hard-core limit, the scattering angle is given by $\chi=\Theta(\sigma-b)[\pi - 2\sin^{-1}(b/\sigma)]$, and we can analytically get the expression $\Omega_{2,2}^{\rm HC}(T)=2\sigma^2 \sqrt{\pi k_{\rm B}T/m}$.
We show the temperature dependence of the Omega integral $\Omega_{2,2}(T)$ in Fig.\ \ref{fig:Omega22}.
We also present the numerical table of the dimensionless Omega integral $\Omega_{2,2}^*(T^*)\equiv \Omega_{2,2}(T)/\Omega_{2,2}^{\rm HC}(T)$ as a function of the dimensionless temperature $T^*\equiv k_{\rm B}T/(k\sigma^2)$.
In the high temperature limit, this integral decreases to zero because $\chi\to \pi/2$ (see Eq.\ \eqref{eq:Q_l}).

%%%%%%%%%%%%%%%%%%%%%%%%%%%%%%
%%%%%%%%%%%%%%%%%%%%%%%%%%%%%%

%%%%%%%%%%%%%%%%%%%%%%%%%%%%%%
\section{Kinetic Theory}\label{sec:kinetic_theory}
Using the information on the scattering angle, let us extend the kinetic theory to this system.
Our starting point is the Boltzmann equation under the simple shear \cite{Hayakawa19, Hayakawa17, Takada20}:
\begin{align}
	&\left(\frac{\partial}{\partial t}-\dot\gamma V_y \frac{\partial}{\partial V_x}\right)f(\bm{V},t)\nonumber\\
	&=\zeta \frac{\partial}{\partial \bm{V}}\cdot \left[\left(\bm{V}+\frac{T_{\rm ex}}{m}\frac{\partial}{\partial \bm{V}}\right)f(\bm{V},t)\right]+J(\bm{V}|f),
\end{align}
where $J(\bm{V}|f)$ is the collision operator given by
\begin{align}
	&J(\bm{V}|f)=\int d\bm{V}_2 \int d\hat{\bm{k}} \Theta(\sigma - b)|\bm{V}_{12}\cdot \hat{\bm{k}}|\nonumber\\
	&\times\left[\sigma_{\rm s}(\chi, V_{12}^\prime)f(\bm{V}_1^\prime,t)f(\bm{V}_2^\prime,t)
	- \sigma_{\rm s}(\chi, V_{12})f(\bm{V}_1,t)f(\bm{V}_2,t)\right].
\end{align}
Here, $\bm{V}_{12}=\bm{V}_1-\bm{V}_2$, $\sigma_{\rm s}(\chi,V)=(b/\sin\chi)|\partial b/\partial \chi|$ is the scattering cross section as a function of the scattering angle $\chi$ and the relative speed $V$.
We also note that $\{\bm{V}_1, \bm{V}_2\}$ and $\{\bm{V}_1^\prime, \bm{V}_2^\prime\}$ are related with each other as
\begin{equation}
	\bm{V}_1^\prime = \bm{V}_1 -\left(\bm{V}_{12}\cdot \hat{\bm{k}}\right)\hat{\bm{k}},\quad
	\bm{V}_2^\prime = \bm{V}_2 +\left(\bm{V}_{12}\cdot \hat{\bm{k}}\right)\hat{\bm{k}}.
\end{equation}
We put $\zeta\propto \sqrt{T_{\rm ex}}$, which means that the drag is determined by the solvent, which is characterized by the external temperature $T_{\rm ex}$.
Here, to characterize the magnitude of the drag force, we define the dimensionless quantity $\xi_{\rm ex}=\sqrt{k_{\rm B}T_{\rm ex}/m}/(\sigma \zeta)$.

For further calculation, we use the Grad approximation to obtain the explicit expressions of the flow curve.
First, we assume that the distribution function is approximately given in terms of Grad's moment method by
\begin{equation}
	f(\bm{V},t)=f_{\rm eq}(\bm{V},t)\exp\left(1+\frac{m}{2nk_{\rm B}^2T^2}P_{\alpha\beta}V_\alpha V_\beta\right),
	\label{eq:Grad}
\end{equation}
with the Maxwellian distribution function
\begin{equation}
	f_{\rm eq}(\bm{V},t)=n\left(\frac{m}{2\pi k_{\rm B}T}\right)^{3/2}\exp\left(-\frac{mV^2}{2k_{\rm B}T}\right),
\end{equation}
where $n$ is the number density of the system.
We can write the second moment of the Boltzmann equation as
\begin{equation}
	\frac{dP_{\alpha\beta}}{dt}+\dot\gamma (\delta_{\alpha x}P_{y\beta}+\delta_{\beta x}P_{y\alpha})
	= -\Lambda_{\alpha\beta} + 2\zeta (nT_{\rm ex}\delta_{\alpha\beta}-P_{\alpha\beta}).
\end{equation}
Here, $\Lambda_{\alpha\beta}$ is defined by
\begin{equation}
	\Lambda_{\alpha\beta} \equiv -m \int d\bm{V}_1 V_{1,\alpha} V_{1,\beta} J(\bm{V}|f),
\end{equation}
and once we adopt the Grad approximation, this quantity becomes
\begin{equation}
	\Lambda_{\alpha\beta}=\nu \left(P_{\alpha\beta}-nk_{\rm B}T\delta_{\alpha\beta}\right),
\end{equation}
(see also Ref.\ \cite{Hayakawa19}).
Here, $\nu$ is defined by
\begin{equation}
	\nu(T)\equiv\frac{8}{5}n\Omega_{2,2}(T).
\end{equation}

The time evolutions of the temperature, the temperature difference, and the shear stress are given by
\begin{align}
	\frac{dT}{dt}&= -\frac{2\dot\gamma}{3nk_{\rm B}}P_{xy} +2\zeta(T_{\rm ex}-T),\label{eq:evol1}\\
	\frac{d\Delta T}{dt}&= -\frac{2\dot\gamma}{nk_{\rm B}}P_{xy}-(\nu+2\zeta)\Delta T,\label{eq:evol2}\\
	\frac{dP_{xy}}{dt}&= \dot\gamma nk_{\rm B} \left(\frac{1}{3}\Delta T - T\right) - (\nu+2\zeta)P_{xy},\label{eq:evol3}
\end{align}
respectively \cite{Hayakawa19, Hayakawa17}.

%%%%%%%%%%%%%%%%%%%%%%%%%%%%%%
\section{Rheology}\label{sec:rheology}

In this section, we investigate the rheology.
First, let us introduce the dimensionless quantities in terms of $m$, $\sigma$, and $\zeta$.
Here, we characterize the stiffness of the particles $k$ as the dimensionless form $k^*\equiv k/(m\zeta^2)$.
As explained in the previous section, the important parameters which characterize the rheology are $T$, $\Delta T$, and $P_{xy}$.
We introduce their dimensionless forms as
\begin{equation}
	\theta\equiv \frac{T}{T_{\rm ex}},\quad
	\Delta\theta \equiv \frac{\Delta T}{T_{\rm ex}},\quad
	\Pi_{xy}^*\equiv \frac{P_{xy}}{nk_{\rm B}T_{\rm ex}},
\end{equation}
respectively.
The dimensionless forms of the shear rate $\dot\gamma$ and the frequency $\nu$ are also, respectively, represented by
\begin{align}
	\dot\gamma^* &\equiv \frac{\dot\gamma}{\zeta},\\
	\nu^* &\equiv \frac{\nu}{\zeta}=\frac{96}{5\sqrt{\pi}}\Omega_{2,2}^* \varphi \xi_{\rm ex}\sqrt{\theta},
\end{align}
where we have introduced the packing fraction $\varphi=(\pi/6)n\sigma^3$.
In the steady state, we can rewrite a set of Eqs.\ \eqref{eq:evol1}--\eqref{eq:evol3} as
\begin{align}
	0&= -\frac{2}{3}\dot\gamma^* \Pi_{xy}^* + 2(1-\theta),\label{eq:evol1_ver2}\\
	0&= -2\dot\gamma^* \Pi_{xy}^* - (\nu^*+2) \Delta\theta,\label{eq:evol2_ver2}\\
	0&= \dot\gamma^* \left(\frac{1}{3}\Delta\theta - \theta\right) - (\nu^*+2)\Pi_{xy}^*.\label{eq:evol3_ver2}
\end{align}
The combination of this treatment in Eq.\ \eqref{eq:evol1_ver2} yields
\begin{equation}
	\Pi_{xy}^*= -\frac{3(\theta-1)}{\dot\gamma^*}.\label{eq:steady_Pxy}
\end{equation}
The temperature difference is determined from Eqs.\ \eqref{eq:evol1_ver2} and \eqref{eq:evol2_ver2} as
\begin{equation}
	\Delta\theta = \frac{6(\theta-1)}{\nu^*+2}.\label{eq:steady_Delta_theta}
\end{equation}
Substituting Eqs.\ \eqref{eq:steady_Pxy} and \eqref{eq:steady_Delta_theta} into Eq.\ \eqref{eq:evol3_ver2}, we obtain
\begin{equation}
	\dot\gamma^*=(\nu^*+2)\sqrt{\frac{3(\theta-1)}{\nu^*\theta +2}}.
\end{equation}
Similarly, the expression of the dimensionless shear viscosity $\eta^*\equiv -\Pi_{xy}^*/\dot\gamma^*$ is given by
\begin{equation}
	\eta^*=\frac{\nu^*\theta+2}{(\nu^*+2)^2}.
\end{equation}
It should be noted that these expressions in the hard-core limit are consistent with those reported in Ref.\ \cite{Hayakawa19}.
These results show that the steady state temperature should be larger than the external temperature $T_{\rm ex}$ because the noise term is kept with the external temperature.
This fact is also reported in Ref.\ \cite{Hayakawa19}.

%%%%%%%%%%%%%%%%%%%%%%%%%%%%%%
\begin{figure}[htbp]
	\centering
	\includegraphics[width=\linewidth]{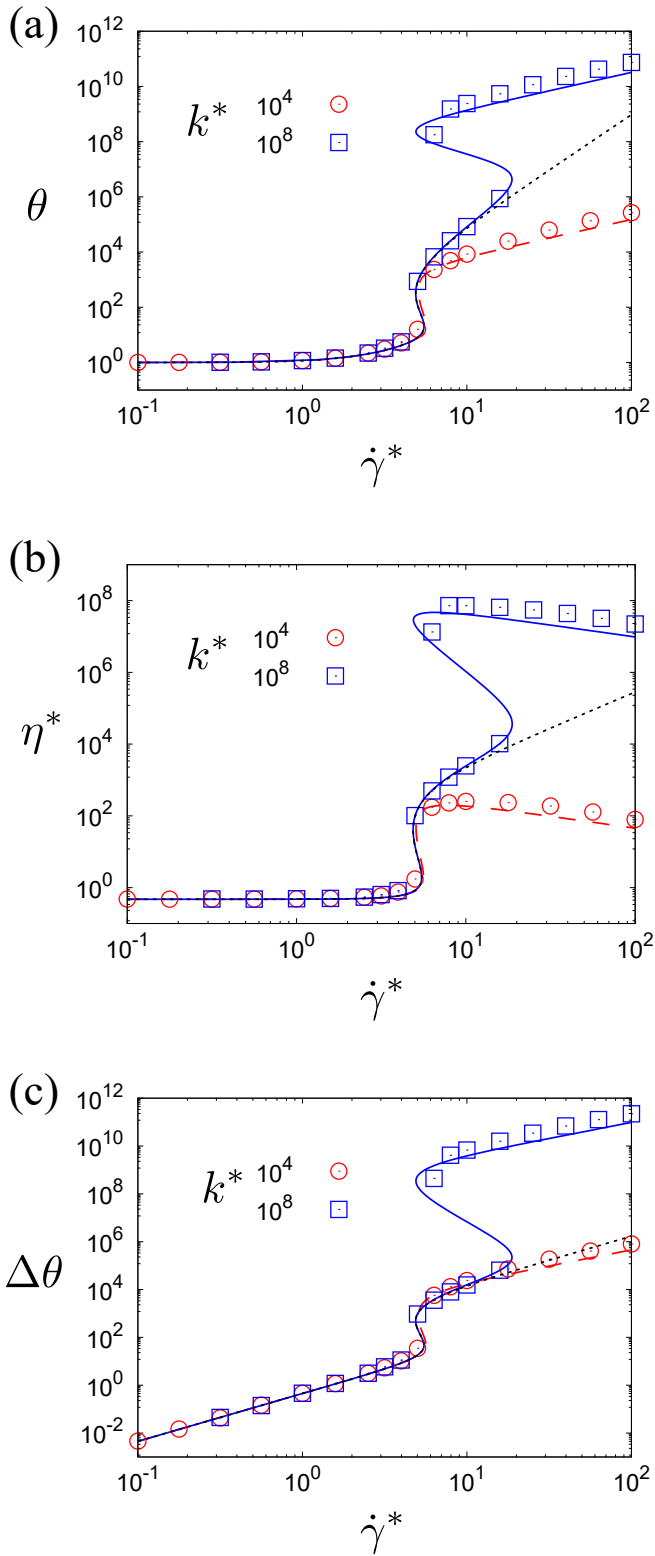}
	\caption{(Color online) Plots of (a) the temperature, (b) the shear viscosity, and (c) the temperature difference for $k^*=1.0\times 10^4\xi_{\rm ex}^2$ (dashed line) and $1.0\times 10^8\xi_{\rm ex}^2$ (solid line) with $\xi_{\rm ex}=1$ and $\varphi=1.0\times10^{-2}$.
	The corresponding simulations results are also plotted ($k^*=1.0\times 10^4\xi_{\rm ex}^2$ (open circles) and $1.0\times 10^8\xi_{\rm ex}^2$ (open squares)).
	The dotted lines represent the theoretical results for the hard-core system \cite{Hayakawa19}.}
	\label{fig:coeff}
\end{figure}
%%%%%%%%%%%%%%%%%%%%%%%%%%%%%%

Figure \ref{fig:coeff} shows the shear rate dependences of the steady temperature, the steady shear viscosity, and the steady temperature difference for $k^*=1.0\times 10^4\xi_{\rm ex}^2$ and $1.0\times 10^8\xi_{\rm ex}^2$ with the magnitude of the external temperature $\xi_{\rm ex}=1$ and the packing fraction $\varphi=1.0\times10^{-2}$.
For $k^*=1.0\times 10^4\xi_{\rm ex}^2$, the flow curves show S-shape in the intermediate shear rate and the shear thinning behavior in the high shear regime, the former of which is also observed those for the hard-core system \cite{Hayakawa19}.
More brilliant change of the flow curve is observed for $k^*=1.0\times 10^8\xi_{\rm ex}^2$.
Here, the S-shape appears twice in the intermediate shear rate.
This is never observed for the hard-core system, which means that this two-step change is originated from the softness of the particles.
The existence of the S-shape directly relates to the discontinuous shear thickening as discussed in Ref.\ \cite{Hayakawa19}, because the abrupt change of the viscosity is observed when we gradually change the shear rate in simulations or experiments.
The shear thinning in the high shear limit is quite different from those for the hard core system, where $\theta\sim (\dot\gamma^*)^4$ and $\eta^*\sim (\dot\gamma^*)^2$.
This is understood by the facts (i) the energy conservation between the energy injection by the shear and the energy dissipation by the background fluid, and (ii) the fact that the collision frequency is proportional to the square of the temperature.
Here, let us consider the condition that the hard-core limit is satisfied.
Because the overlap between particles is not allowed for the hard-core potential, the dimensionless linear spring constant should be sufficiently large.
In addition, the temperature is sufficiently small with respect to the potential energy $k\sigma^2$ as shown in Figs.\ \ref{fig:r_min} and \ref{fig:theta}.
Indeed, the latter condition is not satisfied in the high shear limit in Fig.\ \ref{fig:coeff}, which is the origin of the difference between this system and the hard-core systems.

We also consider the hard-core limit for this system.
For the high temperature regime, the Omega integral $\Omega_{2,2}^*$ and the frequency $\nu^*$ behave as $\Omega_{2,2}^*\sim \theta^{-2}$ and $\nu^*\sim \theta^{-3/2}$, respectively, as shown in Fig.\ \ref{fig:Omega22}.
In this case, the relationship between $\nu^*$, $\nu^*\theta (\sim\theta^{-1/2})$, and $2$ is important.
For our choice of the parameters in Fig.\ \ref{fig:coeff}, $\nu^*\ll 2$ and $\nu^*\theta\gg 2$ are satisfied.
In this case, the asymptotic behaviors become $\theta\sim (\dot\gamma^*)^{4/3}$ and $\eta^*\sim (\dot\gamma^*)^{-2/3}$.
Of course, when the temperature becomes higher and higher, the relationship $\nu^*\theta\ll 2$ should be satisfied, which yields $\theta\sim(\dot\gamma^*)^2$ and $\eta\sim {\rm const.}$

%%%%%%%%%%%%%%%%%%%%%%%%%%%%%%
\section{Simulation}\label{sec:MD}
In this section, we perform the molecular dynamics simulation to check the validity of our kinetic theoretical treatment.
We randomly distribute $N=10^3$ monodisperse particles with the packing fraction $\varphi=1.0\times 10^{-2}$ in the cubic box without any overlaps between particles.
Here, the linear size of the box is $L=(\pi N/(6\varphi))^{1/3}\sigma=37.4\sigma$.
The interparticle potential is given by the harmonic potential \eqref{eq:potential}.
The motion of particles is influenced by the Gaussian thermostat with the temperature $T_{\rm ex}$.
The equation of motion for each particle is given by the Langevin equation \cite{Kawasaki14}
\begin{equation}
	\frac{d\bm{p}_i}{dt}=\bm{F}_i -\zeta \bm{p}_i + \bm{\xi}_i,\label{eq:Langevin}
\end{equation}
with the aid of the Sllod dynamics \cite{Evans84, Evans}.
where $\bm{p}_i\equiv m(\bm{v}_i-\dot\gamma y \hat{e}_x)$ is the peculiar momentum with the unit vector $\hat{e}_x$ in the $x$-direction, $\bm{F}_i$ is the force acting on $i$-th particle, and $\bm{\xi}_i$ is the noise term which satisfies
\begin{align}
	\langle \bm{\xi}_i(t)\rangle &= 0,\\
	\langle \bm{\xi}_i(t) \bm{\xi}_j(t^\prime)\rangle
	&= 2m\zeta T_{\rm ex}\delta_{ij}\bm{1}\delta(t-t^\prime).
\end{align}
Here, the bracket $\langle \cdot \rangle$ mean the average over the noise distribution.
To apply the shear to the system, the Lees-Edwards boundary condition \cite{Lees72} is also adopted in addition to the Sllod dynamics. 

In Fig.\ \ref{fig:coeff}, we plot the simulation results with the corresponding theoretical data.
Quantitatively, both results are consistent with each other, at least, for $\dot\gamma^* \lesssim 10^1$, while the deviations are observed for the high shear regime, which will be discussed in the next section.
However, the qualitative agreements are exhibited in the whole regime because the asymptotic behaviors are same between the simulations and the theory.
It is surprising that the two-step discontinuous changes are observed even by the simulations, which suggests that the existence of the two-step DST is not superficial.

%%%%%%%%%%%%%%%%%%%%%%%%%%%%%%
\section{Discussion}
In this paper, we have performed the MD simulation to check the validate the theoretical treatment.
Although the good agreement is observed for the low and intermediate shear regime, there appears a discrepancy in the high shear regime.
The reason for this discrepancy is still not clear, but we can put some candidates:
First, we have assumed the Grad approximation \eqref{eq:Grad} as the velocity probability function of this system.
Of course, we have only considered the lowest contribution of the deviation from the Maxwellian distribution.
Recently, Saha and Alam \cite{Saha20} have developed the theory in terms of the higher contribution for the hard-core system.
This treatment is also available for this system, but the analysis might be more complicated.
Second, we have only considered the linear term of the shear rate in our treatment, which is partially related to the first reason.
Takada {\it et al.} \cite{Takada20} have recently confirmed that the linear theory cannot capture especially in the vicinity of the S-shape for the hard-core system, and the expansion with respect to the shear rate is needed even when the density is dilute.
This should be important for our system, but this is a future work because the calculation of the collisional contribution is needed.

We also discuss the realization of the setup of this study.
As mentioned in Ref.\ \cite{Takada20}, the situation described by Eq.\ \eqref{eq:Langevin} is realized when the drag force from the background fluid is balanced with the gravity.
This is true only if the system can keep uniform stably, which might be checked by linear or nonlinear stability analysis of the system, but this will be reported in elsewhere.
We also note that there are some experiments in the microgravity conditions \cite{Sack13, Yu20}, where the effect of the gravity becomes much smaller than the surface of the earth, which might be the candidate to perform experiments to check the possibility of the two-step DST.

%%%%%%%%%%%%%%%%%%%%%%%%%%%%%%
\section{Conclusion}
In this paper, we have developed the kinetic theory for inertial suspensions having soft-core potential.
We have derived the explicit expression of the scattering angle as a function of the impact parameter and the relative speed between particles.
Using this angle, we have obtained the steady state temperature and the shear viscosity for the gas-solid suspensions.
The appearance of the S-shape is similar to the hard-core system, but the two-step S-shape is found to occur when we change the stiffness of the particle, which is not observed for the hard-core system.
We have also found that the hard-core limit in the high shear regime is not satisfied because the temperature is not sufficiently small as compared with the potential energy of the particle.

%%%%%%%%%%%%%%%%%%%%%%%%%%%%%%
\section*{Acknowledgement}
One of the authors (ST) thanks to Hisao Hayakawa, Andr\'{e}s Santos, and Vicente Garz\'{o} for their kind and helpful comments during his stay at the University of Extremadura.
This work is partially supported by the Grant-in-Aid of MEXT for Scientific Research (Grant No.\ 20K14428).

%%%%%%%%%%%%%%%%%%%%%%%%%%%%%%
\appendix
\section{Solution of the Quintic Equation}\label{sec:quintic}
In this Appendix, we give a brief explanation of the procedure to obtain the solution of the quintic equation, which is appeared in the denominator of the integrand of Eq.\ \eqref{eq:theta_eq1}.
We rewrite the quintic equation as
\begin{equation}
	\left(u^{*2} + \frac{p+\beta}{2}\right)^2 - \beta\left(u^*-\frac{q}{2\beta}\right)^2=0.
	\label{eq:quintic1}
\end{equation}
This transform is possible when $\beta$ satisfies
\begin{equation}
	\beta (p+\beta)^2 - q^2 = 4r\beta,
\end{equation}
or equivalently,
\begin{equation}
	g(p)\equiv \beta^3 +2p\beta^2 +(p^2-4r)\beta - q^2 = 0.\label{eq:beta_cond}
\end{equation}
First, we try to obtain the explicit form of $\beta$ which satisfies Eq.\ \eqref{eq:beta_cond}.
Let us introduce $\beta_1$ as
\begin{equation}
	\beta_1 \equiv \beta + \frac{2p}{3}.
\end{equation}
Using this quantity, Eq.\ \eqref{eq:beta_cond} is rewritten as
\begin{equation}
	\beta_1^3 + P\beta_1 + Q=0,\label{eq:beta_cond1}
\end{equation}
where the coefficients $P$ and $Q$ are given by
\begin{align}
	P&\equiv -\left(\frac{p^2}{3}+4r\right)
	=-\frac{(2-v^{*2})^2}{3b^{*4}v^{*4}} - \frac{8}{b^{*2}v^{*2}}(<0),\nonumber\\
	Q&\equiv-\frac{2}{27}p^3 -q^2 +\frac{8pr}{3}
	=-\frac{2(2-v^{*2})^3}{81b^{*6}v^{*6}} -\frac{16(1+v^{*2})}{3b^{*2}v^{*2}},
\end{align}
respectively.
Using Cardano's method, one of the solution of Eq.\ \eqref{eq:beta_cond1} is known to be given by
\begin{equation}
	\beta_1 = \left(-\frac{Q}{2}+\sqrt{\Delta}\right)^{1/3} + \left(-\frac{Q}{2}-\sqrt{\Delta}\right)^{1/3}, \label{eq:beta1_Delta_positive}
\end{equation}
with the discriminant \eqref{eq:Delta}.
We note that all the solutions are real for $\Delta\le0$ and one real and two complex solutions exist for $\Delta>0$.
Here, the condition for $\Delta>0$ corresponds to Eq.\ \eqref{eq:v_cond}.

For $\Delta\ge 0$, Eq.\ \eqref{eq:beta_cond} has only one real solution.
Because the root \eqref{eq:beta1_Delta_positive} is real, the real solution of Eq.\ \eqref{eq:beta_cond} is given by
\begin{equation}
	\beta = -\frac{2p}{3} + \left(-\frac{Q}{2}+\sqrt{\Delta}\right)^{1/3} + \left(-\frac{Q}{2}-\sqrt{\Delta}\right)^{1/3}.
	\label{eq:beta_positive}
\end{equation}
It is also noted that this root is always positive because $g(0)=-q^2<0$ is satisfied.

For $\Delta<0$, there appear complex quantities although the final expressions do not include complex values.
Let us introduce two new quantities $A$ and $B$ which satisfy $P=-3A^2$ and $Q=-A^2B$, or equivalently, $A=\sqrt{-P/3}$ and $B=3Q/P$.
Using these quantities, we rewrite Eq.\ \eqref{eq:beta_cond1} as
\begin{equation}
	\beta_1^3 = 3A^2 \beta_1 + A^2 B. \label{eq:beta_cond2}
\end{equation}
Let us put the solution of the equation \eqref{eq:beta_cond2}:
\begin{equation}
	\beta_1 = 2A \cos\alpha.
\end{equation}
If $\alpha$ satisfies
\begin{equation}
	\cos 3\alpha = \frac{B}{2A},\label{eq:alpha_cond}
\end{equation}
one of the solutions of Eq.\ \eqref{eq:beta_cond2} is given by
\begin{equation}
	\beta_1 = 2A\cos\left[\frac{1}{3}\cos^{-1}\frac{B}{2A}\right].
\end{equation}
It is noted that the two other solutions are, similarly, written as
\begin{equation}
	2A\cos\left(\alpha+\frac{2\pi}{3}\right),\quad
	2A\cos\left(\alpha+\frac{4\pi}{3}\right).
\end{equation}
From Eq.\ \eqref{eq:alpha_cond}, the angle $\alpha$ satisfies $0\le 3\alpha\le \pi$, which means that the followings are realized:
\begin{equation}
	\begin{cases}
	2A \cos\alpha>0\\
	2A \cos\alpha\ge 2A\cos\left(\alpha+\frac{2\pi}{3}\right)\\
	2A \cos\alpha\ge 2A\cos\left(\alpha+\frac{4\pi}{3}\right)\\
	2A\cos\left(\alpha+\frac{2\pi}{3}\right)<0
	\end{cases}.
\end{equation}
For the later discussions, we choose the largest one $\beta_1 =2A\cos\alpha$, and therefore, the root of the Eq.\ \eqref{eq:beta_cond} is given by
\begin{equation}
	\beta=-\frac{2p}{3}+2A\cos\left[\frac{1}{3}\cos^{-1}\frac{B}{2A}\right].
	\label{eq:beta_negative}
\end{equation}
We note that this root is also positive because
\begin{align}
	\beta&\ge -\frac{2p}{3}+2\sqrt{\frac{p^2}{9}+\frac{4r}{3}}\nonumber\\
	&=\frac{2p}{3}\left(-1+\sqrt{1+\frac{12r}{p^2}}\right)>0.
\end{align}
Here, we have used the fact that $r$ is always positive from the definition \eqref{eq:def_pqr}.

Using the expressions of $\beta$ given in Eqs.\ \eqref{eq:beta_positive} and \eqref{eq:beta_negative}, let us derive the expression of the solution of Eq.\ \eqref{eq:quintic1}.
Equation \eqref{eq:quintic1} can be rewritten as
\begin{align}
	&\left[\left(u^{*2} + \frac{p+\beta}{2}\right) + \sqrt{\beta}\left(u^*-\frac{q}{2\beta}\right)\right]\nonumber\\
	&\times \left[\left(u^{*2} + \frac{p+\beta}{2}\right) - \sqrt{\beta}\left(u^*-\frac{q}{2\beta}\right)\right]
	=0,
\end{align}
which means that all the solutions are given by
\begin{equation}
	u^*= 
	\begin{cases}
	\displaystyle \frac{-\sqrt{\beta}\pm \sqrt{-\beta - 2p+ \frac{2q}{\sqrt{\beta}}}}{2}\\
	\displaystyle \frac{\sqrt{\beta}\pm \sqrt{-\beta - 2p- \frac{2q}{\sqrt{\beta}}}}{2}
	\end{cases}.
\end{equation}
To check whether these solutions are real or not, let us introduce
\begin{align}
	g_1(u^*)&\equiv u^{*2}+\sqrt{\beta}u^* + \frac{p+\beta}{2} - \frac{q}{2\sqrt{\beta}},\\
	g_2(u^*)&\equiv u^{*2}-\sqrt{\beta}u^* + \frac{p+\beta}{2} + \frac{q}{2\sqrt{\beta}},
\end{align}
and the corresponding discriminants
\begin{align}
	D_1 &=-\beta - 2p + \frac{2q}{\sqrt{\beta}},\\
	D_2 &=-\beta - 2p - \frac{2q}{\sqrt{\beta}}>0,
\end{align}
respectively.
For $g_1(u^*)=0$, because the $x$ coordinate of the vertex is negative, and $g_1(1)>0$, $g_1(u^*)=0$ has no solution which satisfies $u^*>1$.
For $g_2(u^*)=0$, on the other hand, $g_2(u^*)=0$ have two real solutions.
Because $g_2(1)<0$ is satisfied, $g_2(u^*)=0$ have one real solution, and this is given by Eq.\ \eqref{eq:u0}.

%%%%%%%%%%%%%%%%%%%%%%%%%%%%%%

\section{Derivation of the Expression of $\theta$}\label{sec:theta}
In this Appendix, let us show the detailed derivation of the expression $\theta$.
Here, we define $\theta_1$ as
\begin{align}
	\theta_1&\equiv \theta - \sin^{-1}b^* 
	= \int_{1}^{u_0^*}\frac{u^* du^*}{\sqrt{-(u^{*4}+p u^{*2}+qu^*+r)}}.
\end{align}
To calculate this integral, let us introduce a new variable $w$ as
\begin{equation}
	u^* = \frac{\alpha_1+\alpha_2 w}{1+w},
\end{equation}
where $\alpha_1$ and $\alpha_2$ ($\alpha_1\ge \alpha_2$) are the roots of the following equation:
\begin{equation}
	\beta \alpha^2 -2q\alpha -\frac{\beta(p+\beta)}{2}=0,
\end{equation}
that is
\begin{align}
	\alpha_1 &= \frac{q+\sqrt{q^2+2\beta^2(p+\beta)}}{2\beta},\\
	\alpha_2 &= \frac{q-\sqrt{q^2+2\beta^2(p+\beta)}}{2\beta},
\end{align}
respectively.
Using the variable $w$, we can rewrite the integrand as
\begin{align}
	&\frac{u^*du^*}{\sqrt{-(u^{*4}+pu^{*2}+qu^*+r)}}\nonumber\\
	&=-\frac{2\sqrt{A_1A_2}}{(\pm D_1D_2)^{1/4}}\frac{\alpha_1+\alpha_2 w}{1+w}
	\frac{dw}{\sqrt{(A_1^2\mp w^2)(A_2^2-w^2)}},
\end{align}
where the double sign corresponds to the condition: $\beta+2p \lessgtr 2q/\sqrt{\beta}$ (or equivalently, $D_1\gtrless 0$).
We redefine $-w$ as $w$ and we can write
\begin{align}
	\theta_1 &= \frac{2\sqrt{A_1A_2}}{(\pm D_1D_2)^{1/4}}\int_{w_0}^{A_2} \frac{\alpha_1-\alpha_2 w}{1-w}
	\frac{dw}{\sqrt{(A_1^2\mp w^2)(A_2^2-w^2)}}\nonumber\\
	&= \frac{2\sqrt{A_1A_2}}{(\pm D_1D_2)^{1/4}}
	\left[\alpha_2 \tilde{\theta}_1^{(1)}
	+(\alpha_1-\alpha_2)\tilde{\theta}_1^{(2)}\right],
\end{align}
where we have introduced $w_0$ as
\begin{equation}
	w_0 \equiv -\frac{\sqrt{q^2+2\beta^2(p+\beta)}+q-2\beta}{\sqrt{q^2+2\beta^2(p+\beta)}-q+2\beta},
\end{equation}
and we have put
\begin{align}
	\tilde{\theta}_1^{(1)}&\equiv \int_{w_0}^{A_2} \frac{dw}{\sqrt{(A_1^2\mp w^2)(A_2^2-w^2)}},\label{eq:theta1_1}\\
	\tilde{\theta}_1^{(2)}&\equiv \int_{w_0}^{A_2} \frac{dw}{(1-w)\sqrt{(A_1^2\mp w^2)(A_2^2-w^2)}},\label{eq:theta1_2}
\end{align}
respectively.

Let us evaluate $\tilde{\theta}_1^{(1)}$ and $\tilde{\theta}_1^{(2)}$.
For $D_1>0$, we introduce $\varphi$ as $w=A_2 \sin\varphi$, and we rewrite Eq.\ \eqref{eq:theta1_1} as
\begin{align}
	\tilde{\theta}_1^{(1)}
	&=\frac{1}{A_1} \int_{\sin^{-1}(w_0/A_2)}^{\pi/2} \frac{d\varphi}{\sqrt{1-\frac{A_2^2}{A_1^2}\sin^2\varphi}}\nonumber\\
	&=\frac{1}{A_1} \left(\int_0^{\pi/2}-\int_0^{\sin^{-1}(w_0/A_2)} \right)\frac{d\varphi}{\sqrt{1-\frac{A_2^2}{A_1^2}\sin^2\varphi}}\nonumber\\
	&=\frac{1}{A_1}\left[K\left(\frac{A_2^2}{A_1^2}\right) - F\left(\sin^{-1}\frac{w_0}{A_2},\frac{A_2^2}{A_1^2}\right)\right].
\end{align}
Using the similar procedure, Eq.\ \eqref{eq:theta1_2} is rewritten as
\begin{align}
	\tilde{\theta}_1^{(2)}
	&= \frac{1}{A_1}\int_{\sin^{-1}(w_0/A_2)}^{\pi/2} \frac{d\varphi}{(1-A_2 \sin\varphi)\sqrt{1-\frac{A_2^2}{A_1^2}\sin^2\varphi}}\nonumber\\
	&= \frac{1}{A_1}\left[\Pi\left(A_2^2\left| \frac{A_2^2}{A_1^2}\right.\right)-\Pi\left(A_2^2; \sin^{-1}\frac{w_0}{A_2} \left| \frac{A_2^2}{A_1^2}\right.\right)\right]\nonumber\\
	&\hspace{1em}+\frac{1}{\sqrt{(A_1^2-1)(1-A_2^2)}}\tan^{-1}\sqrt{\frac{(A_1^2-1)(A_2^2-w_0^2)}{(1-A_2^2)(A_1^2-w_0^2)}}.
\end{align}

We can follow the similar way for $D_1<0$.
Let us introduce $\varphi$ as $w=A_2\cos\varphi$ and we obtain
\begin{align}
	\tilde{\theta}_1^{(1)}
	&=\frac{1}{\sqrt{A_1^2+A_2^2}}F\left(\cos^{-1}\frac{w_0}{A_2},\frac{A_2^2}{A_1^2+A_2^2}\right),
\end{align}
and
\begin{align}
	\tilde{\theta}_1^{(2)}
	&= \frac{1}{(1-A_2^2)\sqrt{A_1^2+A_2^2}}
	\Pi\left(-\frac{A_2^2}{1-A_2^2};\cos^{-1}\frac{w_0}{A_2} \left|\frac{A_2^2}{A_1^2+A_2^2}\right.\right)\nonumber\\
	&\hspace{1em}+\frac{1}{\sqrt{(A_1^2+1)(1-A_2^2)}} \tan^{-1}\sqrt{\frac{(A_1^2+1)(A_2^2-w_0^2)}{(1-A_2^2)(A_1^2+w_0^2)}},
\end{align}
respectively.

We summarize the expressions of $\theta_1$.
For $D_1\ge 0$, 
\begin{align}
	\theta_1
	&= \frac{2\sqrt{A_1A_2}}{(D_1D_2)^{1/4}}
	\left\{\frac{\alpha_2}{A_1}\left[K\left(\frac{A_2^2}{A_1^2}\right) - F\left(\sin^{-1}\frac{w_0}{A_2},\frac{A_2^2}{A_1^2}\right)\right]\right.\nonumber\\
	&\hspace{1em}\left.+\frac{\alpha_1-\alpha_2}{A_1}\left[\Pi\left(A_2^2\left| \frac{A_2^2}{A_1^2}\right.\right)-\Pi\left(A_2^2; \sin^{-1}\frac{w_0}{A_2} \left| \frac{A_2^2}{A_1^2}\right.\right)\right]\right.\nonumber\\
	&\hspace{1em}\left.+\frac{\alpha_1-\alpha_2}{\sqrt{(A_1^2-1)(1-A_2^2)}}\tan^{-1}\sqrt{\frac{(A_1^2-1)(A_2^2-w_0^2)}{(1-A_2^2)(A_1^2-w_0^2)}}\right\},\label{eq:theta1_positive}
\end{align}
and for $D_1<0$,
\begin{align}
	\theta_1
	&= \frac{2\sqrt{A_1A_2}}{(-D_1D_2)^{1/4}}
	\left\{\frac{\alpha_2}{\sqrt{A_1^2+A_2^2}}F\left(\cos^{-1}\frac{w_0}{A_2},\frac{A_2^2}{A_1^2+A_2^2}\right)\right.\nonumber\\
	&\hspace{1em}\left.
	+\frac{\alpha_1-\alpha_2}{(1-A_2^2)\sqrt{A_1^2+A_2^2}}
		\Pi\left(-\frac{A_2^2}{1-A_2^2};\cos^{-1}\frac{w_0}{A_2} \left|\frac{A_2^2}{A_1^2+A_2^2}\right.\right) \right.\nonumber\\
	&\hspace{1em}\left.+\frac{\alpha_1-\alpha_2}{\sqrt{(A_1^2+1)(1-A_2^2)}}\tan^{-1}\sqrt{\frac{(A_1^2+1)(A_2^2-w_0^2)}{(1-A_2^2)(A_1^2+w_0^2)}}\right\}.\label{eq:theta1_negative}
\end{align}
The coefficients appeared in Eqs.\ \eqref{eq:theta1_positive} and \eqref{eq:theta1_negative} are equivalent to those listed in Table \ref{fig:coeff_list}.

%%%%%%%%%%%%%%%%%%%%%%%

\newpage
\setcounter{equation}{0}
\renewcommand{\theequation}{\arabic{equation}}
\renewcommand{\thefigure}{\arabic{figure}}
\renewcommand{\thetable}{\arabic{table}}
\begin{center}
{\Large Addendum to ``Two-Step Discontinuous Shear Thickening of Dilute Inertial Suspensions Having Soft-Core Potential''}
\end{center}

%%%%%%%%%%%%%%%%%%%%%%%%%%%%%%%%%%%%%%%%
Recently, the rheology of dilute inertial suspensions having soft-core potential is theoretically studied \cite{Sugimoto20}, where collisions are assumed to occur in infinitesimal time.
However, the finite duration of contact must be important in denser systems \cite{Kawasaki14}.
Although we cannot treat the contact duration in the present framework, it is meaningful to obtain its information for future applications.
To this end, we derive the explicit form of the duration time of contact in this Addendum.

%%%%%%%%%%%%%%%%%%%%%%%%%%%%%%%%%%%%%%%%
We briefly explain our model, which is the same as that used in Ref.\ \cite{Sugimoto20}.
We consider the monodisperse particles whose mass and diameter are $m$ and $\sigma$, respectively.
The interparticle force is given by the harmonic potential:
\begin{equation}
	U(r)=\frac{k}{2}\left(\sigma-r\right)^2 \Theta\left(\sigma - r\right),
\end{equation}
where $k$ is the magnitude of the repulsion of the particles and $\Theta(x)$ is the step function.

\begin{figure}[htbp]
	\centering
	\includegraphics[width=0.8\linewidth]{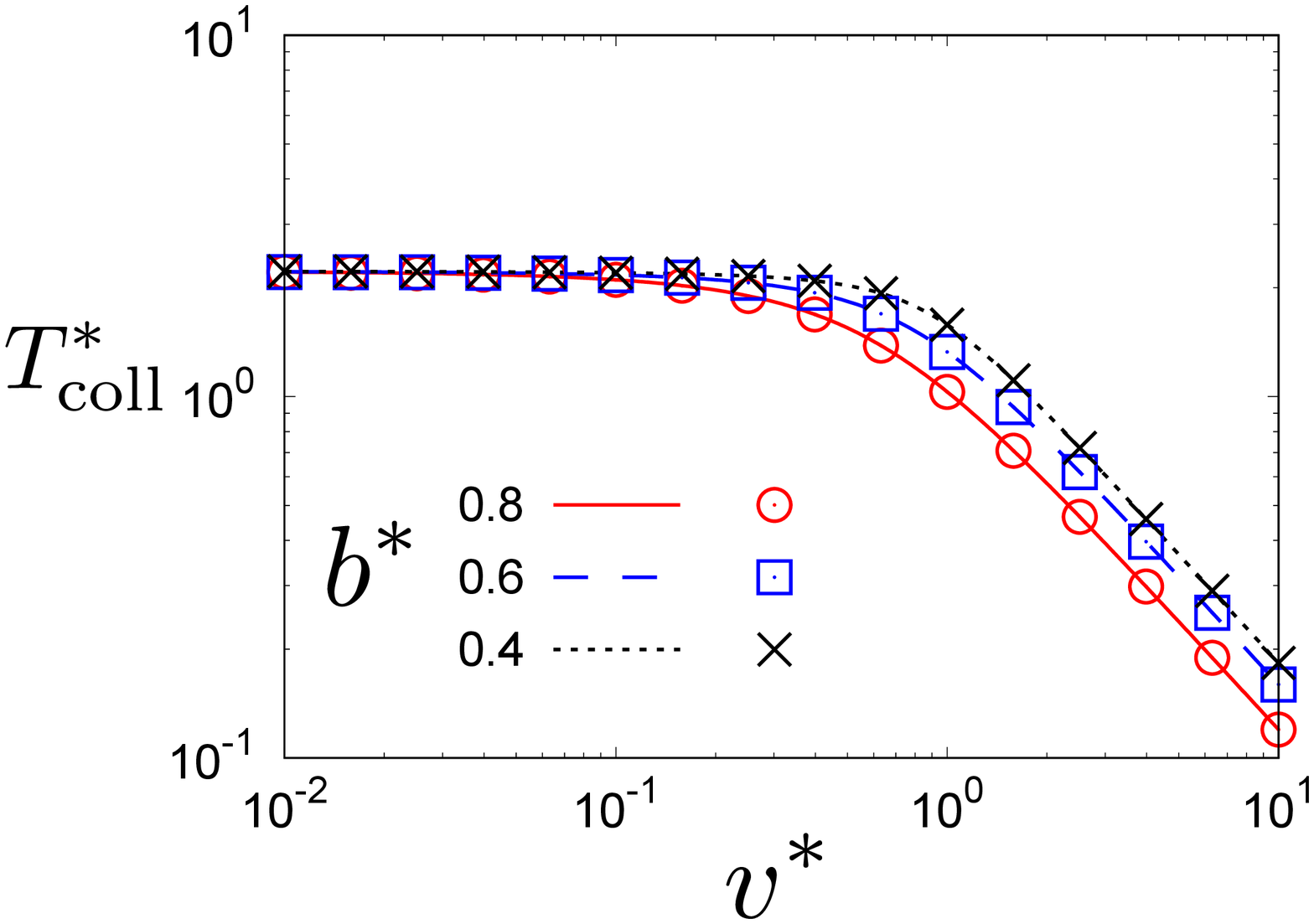}
	\caption{(Color online) Plots of the dimensionless collision duration $T_{\rm coll}^*[\equiv T_{\rm coll}/(k\sigma^2)]$ as a function of the dimensionless relative speed $v^*[\equiv v/(\sigma\sqrt{k/m})]$ for $b^*(\equiv b/\sigma)=0.8$ (red solid line), $0.6$ (blue dashed line), and $0.4$ (black dotted line). 
	The simulation results are also plotted for the same $b^*$.}
	\label{fig:T_coll}
\end{figure}

%%%%%%%%%%%%%%%%%%%%%%%%%%%%%%%%%%%%%%%%
Let us consider the case when two particles collide with each other with the impact parameter $b$ and the relative speed $v$ as shown in Fig.\ 2 of Ref.\ \cite{Sugimoto20}.
The definition of the collision duration is given by \cite{Goldstein}
\begin{align}
	T_{\rm coll} &= 2\int_1^{u_0}\frac{1}{v}\frac{du}{u^2\sqrt{1-b^2u^2-\frac{4}{mv^2}U(1/u)}}\nonumber\\
	&=\frac{2\sigma^2}{bv}\int_1^{u_0^*}\frac{du^*}{u^*\sqrt{-(u^{*4}+pu^{*2}+qu^*+r)}},\label{eq:T}
\end{align}
where the quantities ($p$, $q$, and $r$) appeared in the integrand are listed in Table I of Ref.\ \cite{Sugimoto20, note}.
Here, the integral in Eq.\ \eqref{eq:T} is written in terms of the elliptic integrals \cite{Sugimoto20}.
After similar calculations in Appendix B of Ref.\ \cite{Sugimoto20}, we obtain
\begin{align}
	T_{\rm coll}&=\frac{2\sigma^2}{bv}\left[C_1^{(T)}F(\phi,\nu)+C_2^{(T)}\Pi(a^{(T)};\phi|\nu)\right.\nonumber\\
	&\hspace{6em}\left.+C_3^{(T)}\tan^{-1}\gamma^{(T)}+C_4^{(T)}\right],
	\label{eq:T_coll}
\end{align}
where the expressions of $C_1^{(T)}$, $C_2^{(T)}$, $C_3^{(T)}$, $C_4^{(T)}$, $\gamma^{(T)}$, and $a^{(T)}$ are listed in Table \ref{fig:table}.
We note that the expressions of $\phi$ and $\nu$ are given in Table I of Ref.\ \cite{Sugimoto20}.
%%%%%%%%%%%%%%%%%%%%%%%%%%%%%%
\begin{table}[htbp]
	\centering
	\caption{The expressions of the coefficients appeared in Eq.\ \eqref{eq:T_coll} \cite{note}.}
 	\begin{tabular}{c|cc}
 	\hline
 	 & Condition I & Condition II \\
 	 & $D_1\ge0$ & $D_1<0$ \\ \hline
 	 %%%
 	$\gamma^{(T)}$ & $\sqrt{\frac{\left(\alpha_2^2A_1^2-\alpha_2^2\right)\left(A_2^2-w_0^2\right)}{\left(\alpha_1^2-\alpha_2^2A_2^2\right)\left(A_1^2-w_0^2\right)}}$ & $\sqrt{\frac{\left(\alpha_2^2A_1^2+\alpha_1^2\right)\left(A_2^2-w_0^2\right)}{\left(\alpha_1^2-\alpha_2^2A_2^2\right)\left(A_1^2+w_0^2\right)}}$ \\ 
 	$a^{(T)}$ & $\frac{\alpha_2}{\alpha_1}A_2$ & $-\frac{\alpha_2^2A_2^2}{\alpha_1^2-\alpha_2^2A_2^2}$ \\ 
 	$C_1^{(T)\prime}$ & $-\frac{1}{\alpha_2 A_1}$ & $\frac{1}{\alpha_2 \sqrt{A_1^2+A_2^2}}$ \\ 
 	$C_2^{(T)\prime}$ & $\frac{\alpha_1-\alpha_2}{\alpha_1\alpha_2A_1}$ & $-\frac{\alpha_1 \left(\alpha_1-\alpha_2\right)}{\alpha_2\left(\alpha_1^2-\alpha_2^2A_2^2\right)\sqrt{A_1^2+A_2^2}}$ \\ 
 	$C_3^{(T)\prime}$ & $-\frac{\alpha_1-\alpha_2}{\sqrt{\left(\alpha_2^2A_1^2-\alpha_1^2\right)\left(\alpha_1^2-\alpha_2^2A_2^2\right)}}$ & $-\frac{\alpha_1-\alpha_2}{\sqrt{\left(\alpha_2^2A_1^2+\alpha_1^2\right)\left(\alpha_1^2-\alpha_2^2A_2^2\right)}}$ \\ 
 	$C_4^{(T)\prime}$ & $-C_1^{(T)\prime}K(\nu)-C_2^{(T)\prime}\Pi(a,\nu)$ & $0$ \\ \hline
 	$C_1^{(T)}$ & \multicolumn{2}{c}{$C C_1^{(T)\prime}$} \\ 
 	$C_2^{(T)}$ & \multicolumn{2}{c}{$C C_2^{(T)\prime}$} \\ 
 	$C_3^{(T)}$ & \multicolumn{2}{c}{$C C_3^{(T)\prime}$} \\ 
 	$C_4^{(T)}$ & \multicolumn{2}{c}{$C C_4^{(T)\prime}$} \\ 
 	\hline
	\end{tabular}
	\label{fig:table}
\end{table}
%%%%%%%%%%%%%%%%%%%%%%%%%%%%%%
Figure \ref{fig:T_coll} shows the velocity dependence of the duration time of the collision.
In the high speed regime, the duration time decreases with $T_{\rm coll}\propto 1/v$, which is because the trajectory is straightforward.
On the other hand, the duration time converges to the constant ($\pi/\sqrt{2}\simeq2.22$) in the low speed regime.
This can be easily understood from a solution of the second order differential equation with respect to the distance between two colliding particles.
We also note that this behavior is also validated by the molecular dynamics simulations.

%%%%%%%%%%%%%%%%%%%%%%%%%%%%%%%%%%%%%%%%%%%%%%%%%%%%%%%%%%%%
%%%%%%%%%%%%%%%%%%%%%%%%%%%%%%%%%%%%%%%%%%%%%%%%%%%%%%%%%%%%

%%%%%%%%%%%%%%%%%%%%%%%%%%%%%%%%%%%%%%%%
In this Addendum, we have derived the detailed expression of the collision duration from classical mechanics.
Now, we believe that this information will help us to construct the theory for denser systems, which contains the finite duration of contact.

%%%%%%%%%%%%%%%%%%%%%%%%%%%%%%%%%%%%%%%%
\begin{acknowledgment}
One of the authors (S.T.) thanks Hisao Hayakawa and Kuniyasu Saitoh for their discussions.
This work is supported by the Grant-in-Aid of MEXT for Scientific Research (Grant No.\ JP20K14428).
\end{acknowledgment}
%%%%%%%%%%%%%%%%%%%%%%%%%%%%%%%%%%%%%%%%


\begin{thebibliography}{9}
\bibitem{note}
	$A_2$ in the fifth row of Table I in Ref.\ \cite{Sugimoto20} should read $A_1$.
	Similarly, $C_4^\prime$ for Condition I should read $-C_1^\prime K(\nu)-C_2^\prime\Pi(a,\nu)$.
\end{thebibliography}
\end{document}